\documentclass[review]{elsarticle}
\usepackage{amsmath}
\usepackage{graphicx}

\begin{document}

\begin{frontmatter}
 
\title{Effective two-mode description of a few ultra-cold bosons in a double-well potential}
\author[inst]{Jacek Dobrzyniecki\corref{cor1}}
\cortext[cor1]{Corresponding author}
\ead{dobrzyniecki@ifpan.edu.pl}
\author[inst]{Tomasz Sowi\'nski}

\address[inst]{                  
   Institute of Physics, Polish Academy of Sciences, Aleja Lotnikow 32/46, PL-02668 Warsaw, Poland
  }

\begin{abstract}
We present a construction of an improved two-mode model for modeling the dynamics of interacting ultra-cold bosons confined in a one-dimensional double well trap. Unlike in the typically used two-mode model based on the lowest single-particle eigenstates of the external potential, the improved model uses a basis of properly chosen effective wave functions originating in the many-body model. Accuracy of the improved model is examined and it is shown that within a certain limit of inter-particle interaction strength, the model recovers an exact evolution of the wells' populations much more closely than the traditional two-mode model. 
\end{abstract}

\begin{keyword}
 ultra-cold bosons \sep double-well trap \sep few-body systems
 
 \PACS 67.85.Hj
\end{keyword}

\end{frontmatter}

\maketitle 
\section{Introduction}
Dynamical properties of ultra-cold gases have been enjoying increasing interest since the experimental achievement of the Bose-Einstein condensation in 1995 \cite{bec1,bec2}. Modern experimental methods, including advanced trapping techniques and controlling of mutual interactions, enable experimental investigation of many problems which would previously be considered on a theoretical level only. This opens a whole new research field of strongly correlated systems with potential applications in such fields as quantum computing or quantum simulating of condensed matter problems \cite{Feynman,lewenstein2007,lewenstein2012,sowinski2010}. One example of a widely studied problem in the field is the system of a few particles confined in a double-well potential \cite{smerzi1997,milburn1997,menotti2001,meier2001,salgueiro2007,zollner2008,simon2012,he2012,liu2015,dobrzyniecki2017,tylutki2017}. Such systems have been realized experimentally, and used to study the physics of bosonic condensates with a great effect \cite{andrews1997,shin2004,albiez2005,levy2007,leblanc2011,depaz2014}. 

On a theoretical level, the dynamics of bosons in a double-well system is usually studied in a framework of a simplified two-mode model. The model relies on the assumption that all particles occupying a particular well can be described with a single orbital. Thus, the single-particle basis is limited to two modes, chosen as the lowest-energy wave functions localized in the left and the right well, respectively. They are constructed from the ground and the first excited eigenstate of the single-particle Hamiltonian. In consequence, the dynamics of the bosonic system can be calculated almost straightforwardly \cite{raghavan1999,ostrovskaya2000,ananikian2006,lahaye2009,adhikari2014}. 
Although the model is commonly used, its applicability is essentially limited. The fundamental assumption hidden in this approximation is that the on-site interaction energy is much smaller than the excitation energy needed to reach higher energy levels. It means that the model becomes increasingly inadequate when the interaction strength increases. Additionally, the model completely neglects local inter-particle correlations. In a strong-interaction regime, local multi-particle correlations arise in each well and so the particles in a single site can no longer be adequately described \cite{guo2011,dobrzyniecki2016,sakmann2009,garciamarch2012}. 

For intermediate interactions, some improvement of the two-mode approach can be conceived. In the traditional approach, the shapes of the single-particle wave functions are entirely independent of the interaction strength. By taking into account an influence of inter-particle interactions on the shape of single-particle wave functions, the two-mode description can be improved. Techniques of obtaining improved orbitals through variational and mean-field methods have been studied assuming time-independent \cite{masiello2005,streltsov2006} as well as time-dependent \cite{grond2012,dalton2012,sinatra2000} orbital wave functions.

In this paper we investigate a different, much simpler method, of obtaining an effective time-independent two-mode basis. In our approach the shapes of the basis wave functions emerge naturally after diagonalization of the single-particle density matrix of properly chosen eigenstates of the many-body Hamiltonian. We describe a construction of such an effective basis for a system of two, three, and four interacting bosons in a one-dimensional double-well potential. Then, we examine an accuracy of the resulting two-mode model by comparing its predictions with those obtained by both the exact model and the traditional two-mode model. It is shown that the effective model indeed allows one to extend validity of two-mode approximations to higher interaction strengths. 

\section{The system under study}
\label{sec:systemunderstudy}
We consider a system of $N$ spinless bosons of mass $m$, confined in a one-dimensional double-well potential $V(x)$ and interacting via short-range interactions. We concentrate on systems of $N=2$, $N=3$, and $N=4$ particles, but generalization to larger $N$s, besides numerical complexity, is straightforward. The short-range inter-particle interaction is approximated with a point-like potential $g\delta(x-x')$, where the parameter $g$, related to the $s$-wave scattering length, controls the interaction strength \cite{pethick2008}. Note that in the one-dimensional case the Dirac $\delta$ function is a well-defined self-adjoint Hermitian operator and therefore it does not require any regularization \cite{busch1998}. We focus on repulsive interactions, $g > 0$. Experimentally, a quasi-one-dimensional geometry can be realized by introducing a strong harmonic confinement in two remaining spatial directions. In this way the dynamics in these directions is frozen and particles occupy single ground-states. Consequently, the system becomes effectively one-dimensional.

The many-body Hamiltonian of the system, expressed in the second quantization formalism, has the form: 
\begin{equation}
 \label{HamManyBody}
\hat{\mathcal{H}} = \int\!\! \mathrm{d}x\,\hat{\Psi}^\dagger(x) \mathcal{H}_0 \hat{\Psi}(x)  + \frac{g}{2} \int\!\!\mathrm{d}x\, \hat{\Psi}^\dagger(x) \hat{\Psi}^\dagger(x) \hat{\Psi}(x) \hat{\Psi}(x).
\end{equation}
Here $\hat{\Psi}(x)$ is a bosonic field operator that annihilates a particle at position $x$. The operator fulfills the bosonic commutation relations, $\left[\hat{\Psi}(x),\hat{\Psi}^\dagger(x')\right] = \delta(x-x')$ and $\left[\hat{\Psi}(x),\hat{\Psi}(x')\right] = 0$. The single-particle part of the Hamiltonian has a form
\begin{equation}
\mathcal{H}_0 = -\frac{\hbar^2}{2m} \frac{\mathrm{d}^2}{\mathrm{d}x^2} + V(x).
\end{equation}
We model an external double-well potential $V(x)$ as a combination of a harmonic oscillator potential with frequency $\Omega$, and a Gaussian barrier which separates the central region into two wells: 
\begin{equation}
 V(x) = \hbar\Omega \left[ \frac{m\Omega}{2\hbar} x^2 + \lambda \exp \left(-\frac{m\Omega}{2\hbar} x^2 \right) \right].
\end{equation}
The height of the barrier is directly related to the dimensionless  parameter $\lambda$. In further discussion, we use natural harmonic oscillator units, {\it i.e.}, energy is measured in \(\hbar\Omega\) and length in \(\sqrt{\hbar/m\Omega}\).

The spectrum of $\mathcal{H}_0$ can be found numerically via an exact diagonalization on a dense grid in position representation, giving a set of eigenfunctions $\Phi_i(x)$ and their corresponding eigenenergies $\mathcal{E}_i$ \cite{dobrzyniecki2016}. Following the harmonic oscillator convention, we number the individual states beginning from $i = 0$. For $\lambda = 0$, obviously the well-known harmonic oscillator spectrum is recovered. 

In the analysis of double-well problems, it is usual to adopt a basis of single-particle wave functions $\{\varphi_{Li}(x), \varphi_{Ri}(x)\}$, where the individual states are localized respectively in the left or the right well. These states are constructed as combinations of the odd and even eigenstates of the Hamiltonian: 
  \begin{align}
  \label{eq:lrbasis}
      \varphi_{Ri}(x) &= \frac{1}{\sqrt{2}}(\Phi_{2i}(x) + \Phi_{2i+1}(x)), \nonumber \\
      \varphi_{Li}(x) &= \frac{1}{\sqrt{2}}(\Phi_{2i}(x) - \Phi_{2i+1}(x)).
  \end{align}
Although states  $\{\varphi_{\sigma i}(x)\}$ are not eigenstates of the single-particle Hamiltonian ${\cal H}_0$, they form an orthonormal basis. In this basis the Hamiltonian $\mathcal{H}_0$ has both, diagonal (average energies) and off-diagonal (tunnelings) elements:
\begin{equation}
 \int \varphi^*_{\sigma i}(x) \mathcal{H}_0 \varphi_{\sigma' j}(x) \mathrm{d}x = \delta_{ij}\left[\delta_{\sigma\sigma'} E_i-(1-\delta_{\sigma\sigma'})J_i\right],
\end{equation}
where
\begin{equation}
 E_i = \frac{\mathcal{E}_{2i+1} + \mathcal{E}_{2i}}{2}, J_i = \frac{\mathcal{E}_{2i+1} - \mathcal{E}_{2i}}{2}. 
\end{equation}
The field operator $\hat{\Psi}(x)$ can be decomposed as
\begin{equation}
\label{eq:decomposition}
 \hat{\Psi}(x) = \sum\limits_i \left[ \varphi_{Li}(x) \hat{a}_{Li} + \varphi_{Ri}(x) \hat{a}_{Ri} \right], 
\end{equation}
where $\hat{a}_{\sigma i}$ annihilates a boson in state $\varphi_{\sigma i}(x)$. For numerical purposes the summation index $i$ in the decomposition (\ref{eq:decomposition}) is limited to some cutoff number $i_{max}$. In the case of the system under study, we have verified that $i_{max} = 15$ is sufficient, as the final results do not change significantly for larger $i_{max}$. Therefore in further discussion, we will treat the Hamiltonian with $i_{max} = 15$ as equivalent to the full many-body Hamiltonian (\ref{HamManyBody}). 
By substituting (\ref{eq:decomposition}) into (\ref{HamManyBody}), the Hamiltonian can be written as: 
\begin{align}
\label{eq:manybodyhamiltonian2}
 \hat{\cal H} = &\sum\limits_{i} \left[ E_i (\hat{a}^\dagger_{Li} \hat{a}_{Li} + \hat{a}^\dagger_{Ri} \hat{a}_{Ri}) - J_i (\hat{a}^\dagger_{Li} \hat{a}_{Ri} + \hat{a}^\dagger_{Ri} \hat{a}_{Li}) \right] \nonumber \\
    &+ \frac{1}{2} \sum\limits_{IJKL} U_{IJKL} \hat{a}^\dagger_I \hat{a}^\dagger_J \hat{a}_K \hat{a}_L, 
\end{align}
where the indices $I,J,K,L$ represent double-indices $(\sigma,i)$ identifying single-particle states $\varphi_{\sigma i}(x)$. The interaction terms $U_{IJKL}$ can be calculated as: 
\begin{equation}
 U_{IJKL} = g \int\limits^\infty_{-\infty} \varphi^*_I(x) \varphi^*_J(x) \varphi_K(x) \varphi_L(x) \mathrm{d}x. 
\end{equation}
The spectrum of the Hamiltonian (\ref{eq:manybodyhamiltonian2}) can be calculated numerically. To do so, we express the Hamiltonian in a matrix form in the $N$-particle Fock basis and diagonalize it. Then all properties of the system at any moment can be determined.

Here, our aim is to predict the time evolution of the interacting system of bosons being initially located in the lowest single-particle state of the chosen well. Namely we assume that initially the many-body state of the system is
\begin{equation} \label{IniState}
|\mathtt{ini}\rangle = \frac{1}{\sqrt{N}}\left(\hat{a}^\dagger_{R0}\right)^N|\mathtt{vac}\rangle.
\end{equation}
It means that the state of the system at any later moment $t$ can then be calculated straightforwardly as
\begin{equation}
 \label{eq:timeevolution1}
 |\mathbf{\Psi}(t)\rangle = \sum\limits_k \exp\left( \frac{-i \epsilon_\mathtt{k} t}{\hbar} \right) \langle \mathtt{k} | \mathtt{ini} \rangle |\mathtt{k}\rangle,
\end{equation}
where $|\mathtt{k}\rangle$ and $\epsilon_\mathtt{k}$ are the eigenstates and their corresponding eigenenergies of (\ref{eq:manybodyhamiltonian2}), respectively. It is important to note that $|\mathtt{k}\rangle$ and $\epsilon_\mathtt{k}$ depend directly on interaction strength $g$. However, to simplify the notation we do not write out this dependence explicitly.

\section{Two-mode approximation}
\label{sec:effective}
A two-mode model is a natural approximation of any double-well system. Routinely, it involves choosing $i_{max} = 0$ in the decomposition (\ref{eq:decomposition}), i.e., the single-particle state basis is limited to the two lowest energy states, $\varphi_{L0}(x)$ and $\varphi_{R0}(x)$. Then the field operator $\hat{\Psi}(x)$ can be approximated by 
\begin{equation}
 \label{eq:traditional2Mode}
 \hat{\Psi}(x) \approx \varphi_{L0}(x) \hat{a}_{L0} + \varphi_{R0}(x) \hat{a}_{R0}. 
\end{equation}
By substituting (\ref{eq:traditional2Mode}) into the Hamiltonian (\ref{HamManyBody}), the two-mode many-body Hamiltonian is obtained, and similarly to the full-mode Hamiltonian it can be expressed in matrix form and diagonalized. In consequence, the time evolution can be predicted analogously to  (\ref{eq:timeevolution1}).

In this traditional approximation the dynamics of the non-interacting system ($g = 0$) is reproduced perfectly, since the system remains in the space spanned by two the lowest orbitals, i.e., higher single-particle orbitals can be safely neglected. However, as the interactions increase, couplings to higher orbitals start to play an increasingly important role in the many-body Hamiltonian (\ref{eq:manybodyhamiltonian2}) and a model that neglects these states loses its ability to accurately reproduce the dynamics. 

To overcome this difficulty we propose a modified version of a two-mode approximation taking into account interactions between particles and utilizing our information about the initial state. In this approach the basis wave functions are no longer the solutions of the single-particle Schr{\"o}dinger equation. Rather, the two-mode basis consists a pair of orthogonal wave functions $\phi_A(x), \phi_B(x)$ which are specifically tailored to the system to recover its dynamical properties correctly. These two orbitals depend on interactions, however in the limit of vanishing forces ($g=0$) they can be perfectly obtained as some superpositions of non-interacting orbitals $\varphi_{L0}$ and $\varphi_{R0}$.

Given an effective two-mode basis $\phi_A(x), \phi_B(x)$, it is easy to obtain the approximate dynamics in full analogy to the traditional two-mode approximation. First, we define the annihilation operators $\hat{a}_A$ and $\hat{a}_B$ annihilating bosons in appropriate effective single-particle orbitals $\phi_A(x)$ and $\phi_B(x)$. Then we decompose the field operator as: 
 \begin{equation}
 \label{eq:decompositionAB}
  \hat{\Psi}(x) \approx \phi_A(x) \hat{a}_A + \phi_B(x) \hat{a}_B
 \end{equation}
and substitute this decomposition into the Hamiltonian (\ref{HamManyBody}). An effective two-mode Hamiltonian obtained in this way can be easily diagonalized and an approximate time evolution of the system can be predicted. 

\section{Towards effective orbitals}
 \begin{figure}
 \includegraphics[width=1\linewidth]{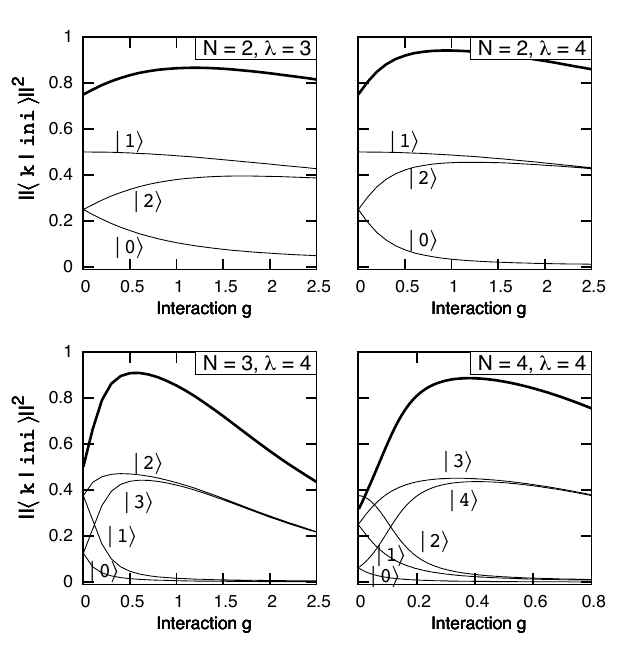}
 \caption{Projection of the initial state $|\mathtt{ini}\rangle$ on consecutive eigenstates of the many-body Hamiltonian \eqref{HamManyBody} as a function of interactions for different numbers of particles (thin black lines). Additionally, the cumulative contribution of the two the most dominant eigenstates $|\mathtt{N}\rangle$ and $|\mathtt{N\!-\!1}\rangle$ is plotted (solid black lines). Note a different scale of interactions in the last plot obtained for $N=4$ particles. Interaction strength $g$ is given in units of $\sqrt{\hbar^3\Omega/m}$.}
 \label{Fig1} 
\end{figure} 

\label{sec:eigenstatebasis}
The most challenging task for an effective two-mode description is to find a proper construction of orbitals $\phi_A(x)$ and $\phi_B(x)$. To make it as good as possible, it is quite obvious that one should take into account not only interactions between particles but also the initial state of the system,  since in principle different initial states are coupled to different orbitals and in consequence they evolve in time completely different. To merge both these requirements, first we decompose an initial state (\ref{IniState}) to the eigenstates of the many-body Hamiltonian (\ref{HamManyBody}). Depending on interactions and the number of particles $N$, the initial state is decomposed to a different number of eigenstates. However, in the non-interacting case ($g=0$), only the lowest $N+1$ eigenstates of the many-body Hamiltonian $|\mathtt{k}\rangle$ (with $\mathtt{k}\in\{0,N\}$) have a nonzero overlap with the initial state, $\langle \mathtt{k} | \mathtt{ini} \rangle$. In this limit all these eigenstates 
can be constructed from two single-particle orbitals $\varphi_{L0}$ and $\varphi_{R0}$ as following
\begin{equation}
|\mathtt{k}\rangle = \frac{1}{\sqrt{k!(N-k)!}}\left(\hat{b}_{+}^\dagger\right)^{N-k}\left(\hat{b}_{-}^\dagger\right)^{k}|\mathtt{vac}\rangle,
\end{equation}
where $\hat{b}_{\pm}=(\hat{a}_{R0}\pm\hat{a}_{L0})/\sqrt{2}$ are symmetric and antisymmetric combinations of annihilation operators in the lowest states of the left and the right well. 
It is quite interesting to note that for non-vanishing interactions the situation becomes in some sense simpler. Although the total number of eigenstates $|\mathtt{k}\rangle$ contributing to the initial state of the system $|\mathtt{ini}\rangle$ increases, only two of them start to dominate in this decomposition. In Fig. \ref{Fig1} we show an overlap of the initial state with consecutive many-body eigenstates $|\mathtt{k}\rangle$ as functions of interactions for different numbers of particles $N$ and chosen depths of the wells $\lambda$. The states are numbered as their counterparts in the limit of vanishing interactions ($g=0$). As it is seen, cumulative contribution of the two selected eigenstates (solid thick lines) remains dominant for a large range of interactions. It means that in this range the initial state $|\mathtt{ini}\rangle$ can be well approximated by proper superposition of only two many-body eigenstates $|\mathtt{N}\rangle$ and $|\mathtt{N\!-\!1}\rangle$. This observation is the first step for our construction. The second is a direct consequence of structural properties of these two many-body eigenstates. For each of these states one can calculate the single-particle density matrix 
\begin{equation} \label{OneDensityMatrix}
\rho^{(k)}(x,x') = \frac{1}{N} \langle \mathtt{k} | \hat{\Psi}^\dagger(x)\hat{\Psi}(x') | \mathtt{k} \rangle,
\end{equation} 
diagonalize it and find its decomposition to the natural single-particle orbitals
\begin{equation}
\rho^{(k)}(x,x') = \sum_i \lambda_i\,  \psi^*_i(x)\psi_i(x').
\end{equation}
For convenience, the orbitals are ordered along their occupations $\lambda_0>\lambda_1>\ldots$. In general, a few the most occupied orbitals $\psi_0$, $\psi_1$, $\psi_2$, \ldots of a single-particle density matrix $\rho^{(k)}$ should be treated as natural candidates for effective orbitals $\phi_A$ and $\phi_B$ carrying information about interactions in the system. From our numerical analysis it follows that two the most important eigenstates $|\mathtt{N}\rangle$ and $|\mathtt{N\!-\!1}\rangle$ with dominant contribution to the initial state share almost the same set of single-particle orbitals $\psi_\lambda$. Moreover, the state $|\mathtt{N}\rangle$ is dominated only by one orbital $\psi_0(x)$, which reproduces the state $\Phi_{0}(x)$ for vanishing interactions. In contrast, the second eigenstate state $|\mathtt{N\!-\!1}\rangle$ is dominated by two orbitals $\tilde\psi_0(x)$ and $\psi_1(x)$ corresponding to single-particle orbitals $\Phi_{0}(x)$ and $\Phi_{1}(x)$, respectively. Of course, single-particle orbitals $\psi_0(x)$ and $\tilde\psi_0(x)$ extracted from the eigenstates $|\mathtt{N}\rangle$ and $|\mathtt{N\!-\!1}\rangle$ are not precisely the same. However, it is possible to deterministically establish two orthogonal orbitals $\phi_A(x)\approx\psi_0(x)\approx\tilde\psi_0(x)$ and $\phi_B(x)\approx\psi_1(x)$ which give the best description of these two many-body eigenstates $|\mathtt{N}\rangle$ and $|\mathtt{N\!-\!1}\rangle$. It is quite obvious that the unique choice of $\phi_A(x)$ and $\phi_B(x)$ does not exist. However, one of the straightforward choices gives the best predictions for the dynamics of the system. The construction is as follows. Since the eigenstate $|\mathtt{N}\rangle$ is dominated only by one orbital, the first orbital $\phi_A(x)$ is simply set as equal to $\psi_0(x)$. This orbital has a natural decomposition into the single-particle basis (\ref{eq:lrbasis}):
\begin{equation}
 \label{eq:Astate}
   \phi_A(x) = \sum_i \left[\lambda_{Li} \varphi_{Li}(x) + \lambda_{Ri} \varphi_{Ri}(x)\right]
 \end{equation}
with some coefficients $\lambda_{\sigma i}$. Subsequently, having these coefficients in hand, one  constructs the second orbital $\phi_B(x)$ as follows:
\begin{equation}
 \label{eq:makingBstate}
   \phi_B(x) = \sum_i (-1)^i \left[\lambda_{Ri} \varphi_{Li}(x) - \lambda_{Li} \varphi_{Ri}(x)\right].
 \end{equation}
These definitions assure automatically that the modes $\phi_A(x)$ and $\phi_B(x)$ are orthogonal and they reduce to the traditional two-mode basis in the limit of vanishing interactions.

We adopted the approach described above to extract effective single-particle orbitals $\phi_A(x)$ and $\phi_B(x)$ in all the cases up to four particles. In each case studied the procedure and conclusions are the same. Therefore, we believe that the method can be adopted also for larger number of particles. However, as seen in Fig.~\ref{Fig1}, for increasing number of particles the range of interactions where two many-body eigenstates dominate in the decomposition of the initial state rapidly decreases (note different ranges of interactions on different plots).

\section{Accuracy of the model}
\label{sec:results}
 \begin{figure}
 \includegraphics[width=0.8\linewidth]{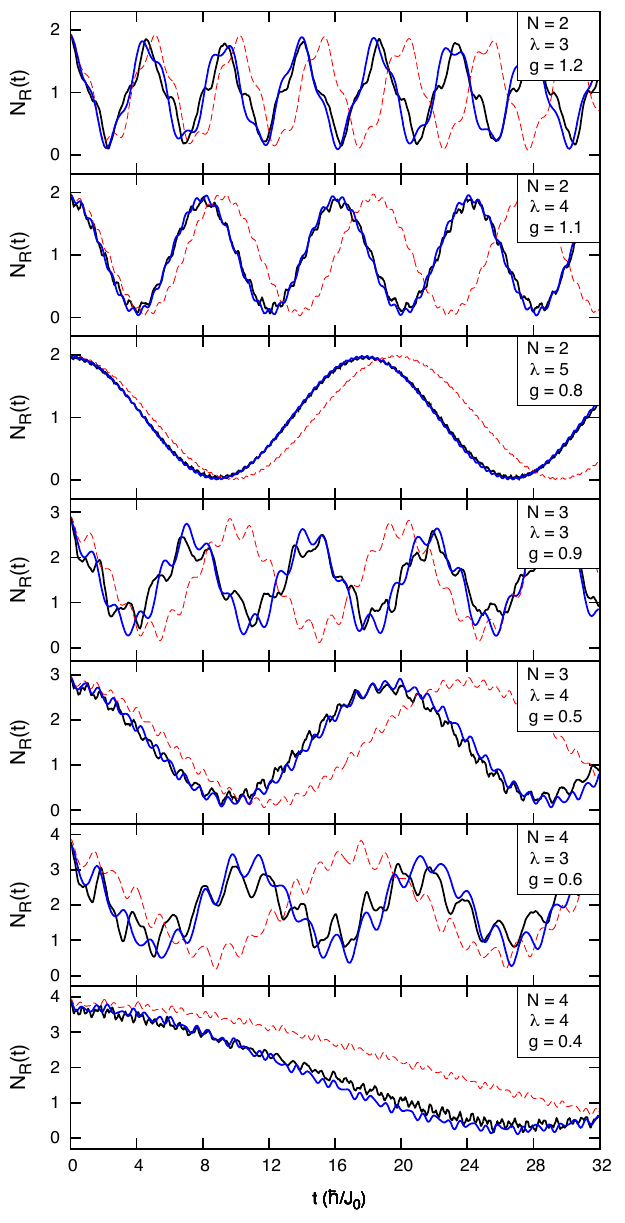}
 \caption{Population of the right well $N_R(t)$ as a function of time predicted by different models studied for different numbers of particles and example parameters of the model. In contrast to the standard two-mode model (red dashed lines), the effective two-mode description (solid blue lines) recovers correctly the results obtained from the full many-body Hamiltonian (solid black). A difference in predictions is clearly visible for longer times. Interaction strength $g$ is given in units of $\sqrt{\hbar^3\Omega/m}$.}
 \label{Fig2}
\end{figure}
First, let us demonstrate a few examples confirming that the effective two-mode model indeed shows improved recovery of the exact dynamics of the system. We focus on the time dependence of the well population which is calculated by integrating a temporal single-particle density over an appropriate region of the space. For the right well it is defined as:
\begin{equation} \label{eq:populationr}
N_{R}(t) = \int_0^\infty \langle \mathbf{\Psi}(t) | \hat{\Psi}^\dagger(x) \hat{\Psi}(x) | \mathbf{\Psi}(t) \rangle \mathrm{d}x.
\end{equation}
The definition for the left well is analogous. 

In Fig. \ref{Fig2} we plot the population of the right well $N_R(t)$ as a function of time as predicted by the full many-body Hamiltonian and by both two-mode models. As it is seen, an evolution of the exact population has a specific oscillatory behavior (solid black lines). In the non-interacting case, oscillations are directly related to the tunneling $J_0$, i.e., the characteristic frequency is equal $\hbar/J_0$. For increasing repulsions, the frequency of the oscillations is modified due to the effects of inter-particle interactions. In particular, tunneling via excited states accelerates a particle flow between the wells. The standard two-mode model, limited to the lowest unperturbed orbitals, is unable to reproduce this effect and in consequence an approximate dynamics generally underestimates the frequency of oscillations (red dashed lines). This causes a growth of the phase shift between the exact and approximate values of $N_R(t)$. On the other hand, as it is seen in Fig. \ref{Fig2}, the effective two-mode model almost exactly reproduces the oscillation frequency and only some small deviations from the exact predictions are visible. 

\begin{figure}
  \includegraphics[width=\linewidth]{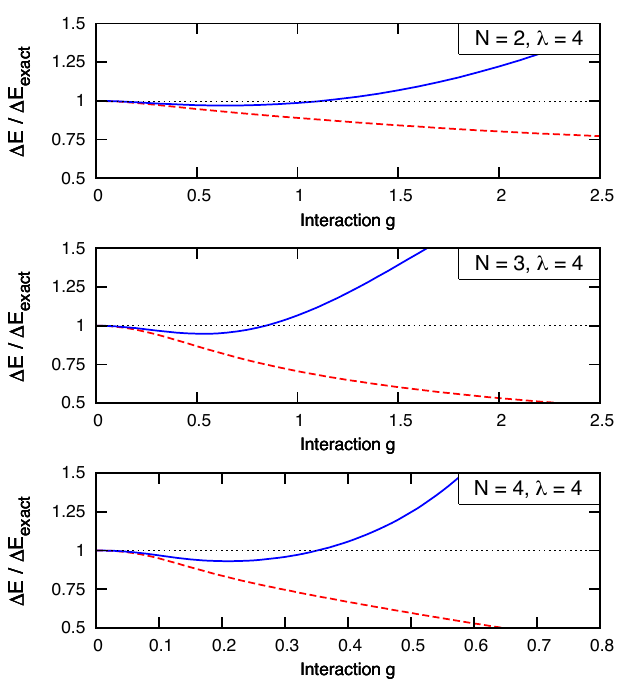}
 \caption{The difference $\Delta E$ between eigenenergies of two the most contributing eigenstates to the initial state calculated in a framework of a chosen two-mode approximation, divided by the same quantity obtained from the exact model $\Delta E_{\mathrm{exact}}$. Red dashed lines correspond to the standard two-mode approximation while solid blue lines to the effective model. Note that in the wide range of interaction strengths $g$ the effective model, in contrast to the standard one, recovers almost perfectly the difference $\Delta E$. In consequence, the effective two-mode description is able to predict oscillations of the population $N_R(t)$ correctly. Interaction strength $g$ is given in units of $\sqrt{\hbar^3\Omega/m}$.}
 \label{Fig3}
\end{figure}

One way to understand sources of the improved accuracy of the effective two-mode model is to analyze the spectrum of the many-body Hamiltonian from the two-mode approximation point of view. In this case the state $|\mathtt{ini}\rangle$ is the only many-body state which describes all particles occupying the right well. Therefore, the frequency of oscillations is directly related to the overlap between the temporal state of the system $|\boldsymbol{\Psi}(t)\rangle$ and the initial state $|\mathtt{ini}\rangle$. Since for the considered initial state, there are two eigenstates of the many-body Hamiltonian $|\mathtt{N}\rangle$ and $|\mathtt{N\!-\!1}\rangle$ which have significant contribution, therefore the expression for time evolution of the state of the system (\ref{eq:timeevolution1}) can be reduced to the sum of two terms only
\begin{align}
 |\mathbf{\Psi}(t)\rangle &\approx \langle \mathtt{N\!-\!1} | \mathtt{ini}\rangle\,\mathrm{e}^{-i\epsilon_\mathtt{N\!-\!1} t/\hbar} |\mathtt{N\!-\!1} \rangle \nonumber \\
&+ \langle \mathtt{N} | \mathtt{ini}\rangle\, \mathrm{e}^{-i\epsilon_\mathtt{N} t/\hbar} | \mathtt{N} \rangle.
\end{align}
In consequence the resulting overlap is simply written as
\begin{equation}
 \lVert \langle \mathtt{ini} | \mathbf{\Psi}(t) \rangle \rVert ^2 \approx C_1 + C_2 \cos{\frac{(\epsilon_\mathtt{N} - \epsilon_\mathtt{N-1})t}{\hbar}},
\end{equation}
with some well established constants $C_1$ and $C_2$. 
This simple analysis shows that the frequency of the dominant Fourier component is related to the difference $\Delta E$ of two eigenenergies of the two the most dominant eigenstates of the Hamiltonian. Depending on the model considered the eigenenergies of the many-body Hamiltonian are different. However, in a wide range of interactions, the difference $\Delta E$ is much closer to the exact value when it is obtained from the effective two-mode description than calculated in the framework of standard approximation. In Fig.~\ref{Fig3} we plot the difference $\Delta E$ divided by its value obtained from the exact model $\Delta E_{\mathrm{exact}}$. As it is seen, this ratio rapidly drops down in the case of the standard two-mode approximation (dashed red lines) while it remains very close to $1$ for a wide range of interactions in the case of the effective model (solid blue lines). In consequence, the frequency of the well population is preserved. This close agreement between the exact and the effective model can be attributed to the effective basis functions which directly take interparticle interactions into account.
\begin{figure}
 \includegraphics[width=\linewidth]{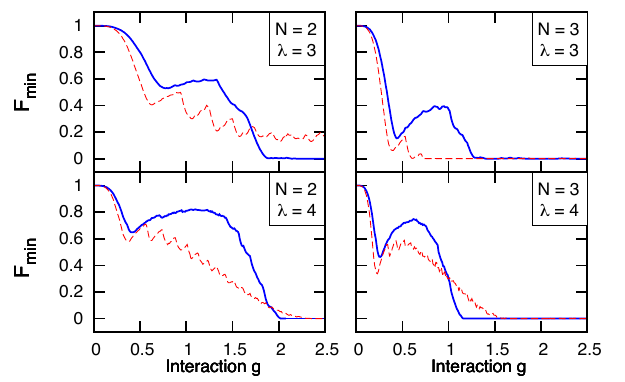}
 \caption{The smallest fidelity $\mathcal{F}_{min}$ achieved by different two-mode models as a function of interactions $g$. The red dashed lines and blue solid lines represent ${\cal F}_\mathrm{min}$ obtained for the traditional two-mode model and the effective two-mode model, respectively. The effective model describes the state of the system much better than the traditional approximation, especially for intermediate interactions. See the main text for details. Interaction strength $g$ is given in units of $\sqrt{\hbar^3\Omega/m}$.}
 \label{Fig4}
\end{figure}

In order to systematically and qualitatively compare accuracies of both two-mode models one should focus not only on the single-particle observables but also on a full quantum many-body state. To make such a comparison possible we introduce a temporal fidelity ${\cal F}(t)=\lVert \langle \mathbf{\Psi}(t) | \psi(t) \rangle \rVert^2$ as a measure of actual accuracy. Here, $|\boldsymbol{\Psi}(t)\rangle$ and $|\psi(t)\rangle$ are the quantum many-body states of the system predicted by the exact model and a chosen two-mode model, respectively. Of course the fidelity defined in this way varies in time. Therefore, we assume that the quality of chosen model is determined by the smallest value of ${\cal F}(t)$ in a chosen time period $t \in (0,T)$. For our purposes we take $T=5\pi\hbar/J_0$, i.e., the period in which the noninteracting system undergoes $5$ oscillations between the wells. In Fig.~\ref{Fig4} we show the smallest fidelity ${\cal F}_\mathrm{min}$ as a function of interactions calculated for the traditional two-mode model (red dashed lines) and for the effective description (solid blue line). This comparison shows that there is a visible improvement in the description of the many-body dynamics over a range of intermediate interaction strengths. However, for strong interactions both models work equally bad. The reason is that any two-mode description has to break down at the moment when strong correlations between particles emerge.

\section{Conclusion}
\label{sec:conclusion}
We present an alternative way to obtain an appropriate two-mode description of the dynamics of a few bosons in a double-well potential. Our approach originates in the decomposition of the initial many-body state in the basis of exact eigenstates of the many-body Hamiltonian. Therefore, it takes into account interactions between particles as well as properties of the initial many-body quantum state. Consequently, the traditional two-mode single-particle basis is replaced by an alternative pair of wave functions, specifically tailored to the problem studied. These wave functions are extracted from the single-particle density matrices of specifically selected eigenstates of the many-body Hamiltonian. 

In consequence, we have shown that for systems of interacting bosons in a double-well potential, the resulting effective two-mode model significantly increases accuracy of the evolution in the range of intermediate interactions where the traditional two-mode model completely fails. The range of interaction strengths for which the model is applicable depends on parameters such as the number of particles, or the height of the barrier between the wells. 

The method presented here relies specifically on the properties of the chosen initial state and, to some extent, it can be generalized to other physical situations. For example, different initial states and larger numbers of particles can be considered. One of the other possible generalizations originates in extending the description to a few the lowest effective orbitals obtained in a very similar way. Although increasing of the number of modes substantially increases the complexity of numerical calculations, a few-mode model seems to be significantly simpler than a full model with many single-particle orbitals taken directly from the non-interacting problem. 

\section{Acknowledgments}  
The authors would like to thank M. Gajda and M. Lewenstein for their fruitful suggestions and questions. This work was partially supported by the (Polish) National Science Center Grant No. 2016/22/E/ST2/00555.

\section*{References}

\bibliography{manuscript}

\end{document}